
\input epsf
\input phyzzx
\overfullrule=0pt
\hsize=6.5truein
\vsize=9.0truein
\voffset=-0.1truein
\hoffset=-0.1truein

%
%

\def\EXP{\hbox{EXP}}
\def\half{{1\over 2}}

\def\IC{{\ \hbox{{\rm I}\kern-.6em\hbox{\bf C}}}}
\def\IR{{\hbox{{\rm I}\kern-.2em\hbox{\rm R}}}}
\def\IZ{{\hbox{{\rm Z}\kern-.4em\hbox{\rm Z}}}}
\def\LOG{\hbox{LOG}}
\def\Mbar{{\cal {\overline M}}}
\def\pa{\partial}
\def\Rbar{{\cal {\overline R}}}
\def\sIR{{\hbox{{\sevenrm I}\kern-.2em\hbox{\sevenrm R}}}}
\def\SC{\hbox{SC}}
\def\Tbar{{\cal {\overline T}}}
\def\Tr{\hbox{Tr}}
\def\Vol{\hbox{Vol}}

%
%
\hyphenation{Min-kow-ski}

\rightline{SU-ITP-94-1}
\rightline{January 1994}
\rightline{hep-th/9401070}

\vfill

%
%
\title{Black Hole Entropy in Canonical Quantum Gravity and
Superstring Theory}

\vfill

%
%
\author{Leonard Susskind\foot{susskind@dormouse.stanford.edu}
and John Uglum\foot{john@dormouse.stanford.edu}}

\vfill

\address{Department of Physics \break Stanford University, Stanford,
CA
94305-4060}

\vfill

%
%
\abstract{\singlespace In this paper the entropy of an eternal
Schwarzschild black hole is studied in the limit of infinite black
hole mass.  The problem is addressed from the point of view of both
canonical quantum gravity and superstring theory.  The entropy per
unit area of a free scalar field propagating in a fixed black hole
background is shown to be quadratically divergent near the horizon.
It is shown that such quantum corrections to the entropy per unit
area are equivalent to the quantum corrections to the gravitational
coupling.  Unlike field theory, superstring theory provides a set of
identifiable configurations which give rise to the classical
contribution to the entropy per unit area.  These configurations can
be understood as open superstrings with both ends attached to the
horizon.  The entropy per unit area is shown to be finite to all
orders in superstring perturbation theory.  The importance of these
conclusions to the resolution of the problem of black hole
information loss is reiterated.}

%
%
PACS categories: 04.70.Dy, 04.60.Ds, 11.25.Mj, 97.60.Lf
\vfill\endpage

%
%

\REF\Lennyone{L.~Susskind, {\it Some Speculations About Black Hole
Entropy in String Theory,} Rutgers University preprint RU-93-44,
August 1993, hep-th/9309145.}

\REF\Unruh{W.~G.~Unruh \journal Phys. Rev. & D14 (76) 870.}

\REF\tHooft{G.~'t~Hooft \journal Nucl. Phys. & B256 (85) 727.}

\REF\Srednicki{M.~Srednicki \journal Phys. Rev. Lett. & 71 (93) 666.}

\REF\Gibhawk{G.~W.~Gibbons and S.~W.~Hawking \journal Phys. Rev. &
D15 (77) 2752.}

\REF\Claudio{M.~Ba\~nados, C.~Teitelboim, and J.~Zanelli \journal
Phys. Rev. Lett. & 72 (94) 957; \hfil\break
S.~Carlip and C.~Teitelboim, {\it The Off Shell Black Hole,} preprint
IASSNS-HEP-93-84, November 1993, gr-qc/9312002.}

\REF\CGHS{C.~G.~Callan, S.~B.~Giddings, J.~A.~Harvey, and
A.~Strominger \journal  Phys. Rev. & D45 (92) 1005.}

\REF\Polchinski{J.~Polchinski \journal Commun. Math. Phys. & 104 (86)
37.}

\REF\Junone{J.~Liu and J.~Polchinski \journal Phys. Lett. & B203 (88)
39.}

\REF\Tseyone{A.~A.~Tseytlin \journal Phys. Lett. & B208 (88) 221.}

\REF\GSWtwo{M.~B.~Green, J.~H.~Schwarz, and E.~Witten, {\it
Superstring Theory, Vol. 2}, Cambridge University Press, 1987.}

\REF\Vadim{V.~Kaplunovsky, private communication.}

\REF\STU{L.~Susskind, L.~Thorlacius, and J.~Uglum \journal Phys. Rev.
& D48 (93) 3743.}

%
%

%
%
\chapter{Introduction}

In conventional statistical mechanics, the entropy of a system
originates in the counting of quantum states which are
macroscopically indistinguishable.  The Bekenstein-Hawking entropy
associated with a black hole of mass $M \gg M_{Planck}$ has never
been explained as a consequence of such state counting.  In fact, as
will be shown, in the theory of quantum fields propagating in a fixed
black hole background, the entropy stored in quanta near the horizon
is divergent.  In other words, an infinite number of macroscopically
indistinguishable states of the quantum field exist arbitrarily close
to the horizon.  This is in obvious contradiction with the finiteness
of the Bekenstein-Hawking entropy.  This conflict is at the root of
the information paradoxes of Hawking.  As an example, consider a
process in which information-carrying photons are dropped into a
black hole at an average rate which compensates the Hawking
evaporation.  If one studies this process in the usual approximation
of quantum fields in a fixed black hole background, one discovers a
contradiction.  After a long while, the information stored by photons
near the horizon is much larger than that permitted by the
Bekenstein-Hawking entropy.  Thus, it is claimed that information is
lost.  The real problem is that there is a conflict between the
entropy defined by state counting (which is infinite) and by black
hole thermodynamics (which is finite).  In this paper it will be
shown that the entropy divergences obtained by state counting are
closely related to conventional ultraviolet divergences of canonical
quantum gravity, and therefore that the information puzzles are part
of the problem of the nonrenormalizability of the theory.

On the other hand, string theory is an ultraviolet finite theory of
gravity.  In an earlier paper, it was speculated by one of us that
string theory resolves the puzzles behind the Bekenstein-Hawking
entropy, both by producing a finite entropy and providing an explicit
counting of states [\Lennyone ].  In this paper we provide strong
evidence to support these speculations.  We will show that the
counting of string states near a horizon gives a finite entropy which
agrees with the usual Bekenstein-Hawking result
$$
\sigma_{BH} = {{4\pi M^2} \over G} = {A \over {4G}} \>.
\eqn\BHent
$$
Furthermore, we will be able to identify the string configurations
which lead to this result.  Thus, in string theory there is no way to
hide information in excess of $\sigma_{BH}$ near the horizon.  It
must be reemitted in the radiation.

We will begin by considering the entropy of an eternal Schwarzschild
black hole using canonical quantum gravity.  Using the Euclidean
functional integral formulation of the partition function for
canonical quantum gravity coupled to matter, we will show that the
Bekenstein-Hawking formula \BHent\ is a general result, but that the
gravitational coupling $G$ appearing in equation \BHent\ is the
renormalized gravitational coupling.  In particular, this means that
the divergences in the entropy due to matter fields are the same
divergences one must deal with when trying to renormalize the theory.
 This shows that the question of the finiteness of the entropy is
inextricably intertwined with the renormalization of the
gravitational coupling, and therefore cannot be understood without a
complete knowledge of the ultraviolet behavior of the theory.  We
then show that in its perturbative formulation, the theory of
superstrings propagating in an eternal black hole background gives
rise to a completely finite entropy of the form \BHent , because the
renormalization of $G$ in superstring theory is finite.  We then
demonstrate that the entropy arises from counting states of open
strings with both ends attached to the horizon.

Even before attempting any calculations, however, a problem arises:
to perform a careful statistical mechanical computation for a black
hole with a temperature $\tau$, the black hole must be in thermal
equilibrium with a radiation bath, and this radiation permeates all
of spacetime.  Therefore, one should expect a divergence,
proportional to the volume of space, in any extensive quantities,
such as the Helmholtz free energy.  This problem can be avoided by
considering the limit of an infinitely massive black hole, for which
the Hawking-Unruh temperature is zero.  The resulting geometry
outside the event horizon is described by the Rindler metric.
Beginning with the Schwarzschild metric
$$
g = - \bigl ( 1 - {{2GM} \over r} \bigr ) dt \otimes dt
+ \bigl ( 1 - {{2GM} \over r} {\bigr )}^{-1} dr \otimes dr
+ r^2 d\Omega_2 \>,
\eqn\gSchwarz
$$
define new coordinates $T$ and $s$ by
$$
\eqalign{T &= {t \over {4GM}} \cr
s &= \sqrt{8GM(r-2GM)} \>, \cr}
\eqn\Tands
$$
so that $g$ may be written
$$
g = -s^2 \bigl ( 1 + {{s^2} \over {16 G^2 M^2}}
{\bigr )}^{-1} dT \otimes dT + \bigl ( 1 + {{s^2}
\over {16 G^2 M^2}} \bigr ) ds \otimes ds +
4 G^2 M^2 \bigl ( 1 + {{s^2} \over {16 G^2 M^2}}
{\bigr )}^2 d\Omega_2 \>.
\eqn\gSnew
$$
Taking the limit $M \rightarrow \infty$, the spherical horizon
surface becomes planar, and equation \gSnew\ becomes the Rindler
metric
$$
g = -s^2 dT \otimes dT + ds \otimes ds + dx^2 \otimes dx^2
+ dx^3 \otimes dx^3 \>.
\eqn\Lorinmet
$$
In order to regulate divergences coming from the infinite area of the
Rindler horizon, we shall put in an infrared cutoff by defining $x^2,
x^3 \in \bigl [ -{L \over 2}, {L \over 2} \bigr ]$, and then impose
suitable boundary conditions on the fields.  With this procedure, the
area $A$ of the horizon is simply $L^2$.  The quantities of interest,
such as the entropy per unit area, remain well defined in the limit
$L \rightarrow \infty$.  In addition, the near horizon temperature
also remains finite.

Rindler space ${\cal R}$ may be viewed as the wedge $x^1 \geq |x^0|$
of Minkowski space.  A Rindler observer at constant $s$ corresponds
to a uniformly accelerated observer in Minkowski space.  From the
point of view of such an observer, Rindler space is causally
complete, because the surface $x^0 = -x^1$ corresponds to $T =
-\infty$.  Signals which pass into Rindler space from the region
beneath this surface are viewed as initial data.  Now consider a
system in the Minkowki space vacuum $\ket{0}_M$.  A Rindler observer,
who can sample only the Rindler wedge, must trace over all degrees of
freedom outside Rindler space.  He will therefore view the Minkowski
vacuum as a mixed state.  Remarkably, the density operator for the
Rindler observer is [\Unruh ]
$$
\rho_R = {{\EXP ( - 2\pi H_R )} \over Z} \>,
\eqn\Rindrho
$$
which corresponds to a thermal ensemble at the Rindler temperature
$\tau_R = {1 \over {2\pi}}$.  The Rindler Hamiltonian $H_R$ is given
in terms of the Minkowski space energy momentum tensor as
$$
H_R = \int_{\cal R} d^2 x ds s (T_M)^{00} \>,
\eqn\PRind
$$
where the integral is evaluated on the surface $T=0$.

%
%
\chapter{Statistical mechanics of a scalar field in Rindler space}

In statistical mechanics, the entropy of a system is defined to be
$$
\sigma = -\Tr \bigl ( \rho \LOG ( \rho) \bigr ) \>,
\eqn\entropy
$$
where $\rho$ is the density operator for which equation \entropy\ is
maximized, subject to constraints.  In the microcanonical ensemble,
in which the energy $E$ of the system is held fixed, $\rho$ is given
by $1/N$ times the projection operator for the eigenspace
corresponding to $E$, where $N$ is the dimensionality of this
eigenspace.  The entropy is $\sigma = \log (N)$, the logarithm of the
dimension of the allowed space of states, and represents a counting
of allowed states.  The effect of a more general density operator may
be to weight states differently, but the entropy can still be
interpreted as the logarithm of an effective dimension of a space of
states with non-negligible probability.  For example, a system in
thermal equilibrium with a heat reservoir at temperature $\tau$ is
described by the density operator
$$
\rho = {{\EXP (-\beta H)} \over Z}
\eqn\rhoeqn
$$
where $\beta = 1/\tau$ is the inverse temperature and $H$ is the
Hamiltonian.  The quantity $Z$ is the partition function, and is
defined by
$$
Z = \Tr \bigl ( \EXP (-\beta H) \bigr ) = \exp (-\beta F) \>,
\eqn\pfone
$$
where $F$ is the Helmholtz free energy.  It is easily shown that the
entropy \entropy\ can be obtained from the Helmholtz free energy as
$$
\sigma = \beta^2 {{\pa F} \over {\pa \beta}} \>.
\eqn\entropytwo
$$

The equation \pfone\ for the partition function applies to quantum
theories of fields propagating on a fixed Lorentzian spacetime
manifold ${\cal M}$, described by a metric $g$.  As long as ${\cal
M}$ is stationary, so that thermal equilibrium is a well-defined
concept, one can select a time-like Killing vector $(\pa / \pa x^0)$
and calculate the partition function.

Consider now a scalar field propagating in Rindler space ${\cal R}$.
The action describing the field is
$$
S[\phi] = - \half \int_{\cal R} {\bf  \epsilon}_g \bigl (
(\nabla \phi )^2 + m^2 \phi^2 \bigr ) \>,
\eqn\phiact
$$
where the metric $g$ is given by equation \Lorinmet\ and ${\bf
\epsilon}_g =  \sqrt{ -\det (g_{\mu \nu})} \, dT \wedge ds \wedge
dx^2 \wedge dx^3$ is the volume form corresponding to $g$.  The field
equation is the usual Klein-Gordon equation
$$
(\nabla^2 - m^2) \phi = \bigl ( - {1 \over s^2}{\pa^2 \over
{\pa T^2}} + {\pa^2 \over {\pa s^2}} + {1 \over s}{\pa \over
{\pa s}} + {\pa^2 \over {(\pa x^2)^2}} + {\pa^2 \over {(\pa
x^3)^2}} - m^2 \bigr ) \phi = 0 \>.
\eqn\sceom
$$
$\phi$ may be expanded as
$$
\phi = U(s) \exp (-i\omega T + i {\vec k} \cdot
{\vec x}_{\perp} ) \>,
\eqn\phiexpand
$$
where ${\vec x}_{\perp} = (x^2, x^3)$.  Imposing periodic boundary
conditions in $x^2$ and $x^3$ requires that ${\vec k} = {{2 \pi}
\over L} (n^2, n^3)$, where $(n^2, n^3) \in \IZ^2$.  Defining $\xi =
\sqrt{{\vec k}^2 + m^2}$, the equation for $U$ may be written
$$
\bigl ( {d \over {ds}} (s {d \over {ds}}) - {{(i\omega)^2}
\over s} \bigr ) U(s) = \xi^2 s U(s) \>,
\eqn\Ueqn
$$
which is the modified Bessel equation of order $i\omega$ in
eigenvalue form.  The most general solution of this equation is
$$
U = A I_{i\omega}(\xi s) + B K_{i\omega}(\xi s) \>.
\eqn\Usol
$$
The eigenfunctions and eigenvalues are determined by imposing
boundary conditions.  The unboundedness of $I_{i\omega}(x)$ as $x
\rightarrow \infty$ requires $A=0$. To regulate the theory, we demand
that $U$ vanish at $s = \varepsilon$, where $\varepsilon$ is close to
zero.  Then the eigenvalues are the solutions of the equation
$$
K_{i\omega}(\xi \varepsilon) = 0 \>.
\eqn\eigfreqeqn
$$
Note that equation \eigfreqeqn\ should be viewed as an equation for
the Rindler frequencies $\omega$.  The solutions of this equation can
be approximated as follows.  Define $x = \log (\xi s)$ and $E = \half
\omega^2$, so that equation \Ueqn\ can be rewritten as a time
independent Schr\"odinger equation
$$
\bigr ( - \half {d^2 \over {d x^2}} + \half e^{2x} \bigr )
U(x) = E U
\eqn\Ueqntwo
$$
for a particle of mass $m = 1$ moving in the potential
$$
V(x) = \cases{\half e^{2x}, &if $x > \log (\xi
\varepsilon)$; \cr \infty, &otherwise. \cr}
\eqn\pot
$$
The eigenvalues $E$ can be approximated using the WKB method.  The
turning points of the classical motion occur when $V = E$, or when $x
\in \{ \log (\xi \varepsilon ), \log (\omega) \}$.  The WKB
quantization condition is
$$
n \pi = \int_{\log (\xi \varepsilon)}^{\log (\omega)} dx
\sqrt{2(E - V(x))} \>.
\eqn\quantcond
$$
This integral can be calculated, and leads to the result
$$
n = {\omega \over {2 \pi}} \Bigl [ \log \Bigl ( {{1 + \sqrt{1
- (\xi \varepsilon / \omega)^2}} \over {1 - \sqrt{1 - (\xi
\varepsilon / \omega)^2}}} \Bigr ) - 2\sqrt{1 - (\xi
\varepsilon / \omega)^2} \Bigr ] \>,
\eqn\neqn
$$
which is an implicit equation for the frequencies $\omega$.
Requiring that the square root be real gives the condition $\omega
\ge \xi \varepsilon$.  The eigenfrequencies depend on both the
quantum number $n$ and the wave vector ${\vec k}$, and are denoted by
$\omega_n({\vec k})$.

The partition function for a single mode, labeled by the quantum
numbers $n$ and ${\vec k}$, is given by
$$
Z(\beta; n, {\vec k}) = \sum_{m=0}^{\infty} e^{-m \beta
\omega_n({\vec k})}
= (1 - e^{-\beta \omega_n({\vec k})})^{-1} \>.
\eqn\Zmode
$$
Since the modes are independent, the total partition function is
$$
Z(\beta) = \prod_{n, {\vec k}} Z(\beta; n, {\vec k}) = \exp
(-\beta F(\beta)) \>,
\eqn\Zall
$$
and the Helmholtz free energy is
$$
F(\beta) = - {1 \over \beta} \sum_{n, {\vec k}} \log (Z(\beta;
n, {\vec k})) \>.
\eqn\frengy
$$
Approximating the sums by integrals, equation \frengy\ becomes
$$
F(\beta) = {L^2 \over \beta} \int_{\sIR^2} {{d^2 k} \over
{(2\pi)^2}} \int_{\xi \varepsilon}^{\infty} d\omega {{dn}
\over {d\omega}} \log (1 - e^{-\beta \omega}) \>.
\eqn\Fone
$$
Differentiating equation \neqn\ gives the density of levels
$$
{{dn} \over {d\omega}} = {1 \over {2\pi}} \log \Bigl ( {{1 +
\sqrt{1 - (\xi \varepsilon / \omega)^2}} \over {1 - \sqrt{1
- (\xi \varepsilon / \omega)^2}}} \Bigr ) \>,
\eqn\dndw
$$
and changing the orders of integration, one obtains the expression
$$
F(\beta) = {A \over {(2\pi)^2 \beta}} \int_{\varepsilon
m}^{\infty} d\omega \log (1 - e^{-\beta \omega})
\int_0^{\sqrt{(\omega / \varepsilon)^2 - m^2}} dk k \log \Bigl
( {{1 + \sqrt{1 - (\varepsilon / \omega)^2 (k^2 + m^2)}}
\over {1 - \sqrt{1 - (\varepsilon / \omega)^2 (k^2 + m^2)}}}
\Bigr ) \>,
\eqn\Ftwo
$$
where $L^2 = A$ is the area of the horizon.  After performing the
integral over $k$, equation \Ftwo\ becomes
$$
F(\beta) = {A \over {(2\pi)^2 \beta}} \int_{\varepsilon
m}^{\infty} d\omega \log (1 - e^{-\beta \omega}) \Bigl [
(\omega / \varepsilon)^2 \sqrt{1 - (\varepsilon m / \omega)^2}
+ {m^2 \over 2} \log \Bigl ( {{1 - \sqrt{1 - (\varepsilon m /
\omega)^2}} \over {1 + \sqrt{1 - (\varepsilon m / \omega)^2}}}
\Bigr ) \Bigr ] \>.
\eqn\Fthree
$$
Expanding in powers of the field mass $m$, the leading term is
$$
F(\beta) =  {A \over {(2\pi \varepsilon)^2 \beta}} \int_0^{\infty}
d\omega \omega^2 \log (1 - e^{-\beta \omega}) \>,
\eqn\Ffour
$$
which is integrated to yield
$$
F(\beta) = -{{\pi^2 A} \over {180 \varepsilon^2 \beta^4}} \>.
\eqn\Ffive
$$

The resulting entropy, evaluated at the Rindler temperature $\tau_R =
1 / 2\pi$, is
$$
\sigma_{\phi} = {A \over {360 \pi \varepsilon^2}} \>.
\eqn\phient
$$
The entropy of the $\phi$ field diverges quadratically with the
cutoff $\epsilon$.  This divergence is proportional to the area of
the horizon, however, and only the numerical coefficient depends on
the cutoff procedure.  The nature of this divergent entropy can be
understood in terms of the infinite gravitational redshift between
the horizon and infinity.  Any field mode with finite frequency at
the horizon must have vanishing frequency at infinity.  Therefore,
the dimension of the space of states with arbitrarily small energy is
infinite.  The regulator $\varepsilon$ is an ultraviolet regulator,
which has the effect of cutting off the sum over states.  The result
\phient\ agrees with the entropy of a scalar field propagating
outside a finite mass black hole as calculated by 't Hooft [\tHooft
].

Because the thermal density operator in Rindler space can be obtained
from the Minkowski vacuum by tracing over the degrees of freedom
outside the Rindler wedge, the above calculation is related to the
calculation performed by Srednicki [\Srednicki ].  Srednicki
calculated the entropy resulting from tracing over the degrees of
freedom of a scalar field contained within a spherical cavity when
the field is in the Minkowski vacuum state.  In the limit of a large
sphere, the entropy per unit area calculated in [\Srednicki ] should
agree with that in Rindler space.  The result \phient\ is in
qualitative agreement with the results in [\Srednicki ], although the
numerical coefficient of the quadratically divergent entropy is
substantially different.  This can be attributed to the sensitive
dependence of quadratic divergences on the regulator method.

%
%
\chapter{The Bekenstein-Hawking entropy and the renormalization of
$G$}

The divergent entropy of the scalar field obtained in Chapter 2 is
certainly nonsensical, and an obvious flaw in the calculation is that
the backreaction of the gravitational field has been neglected.
Since field states with arbitrarily high energy were included in the
partition function, one should expect modifications to the background
geometry.  The argument may be raised, therefore, that a calculation
which includes this backreaction may very well lead to a finite
entropy.  In what follows this is addressed.  We will calculate the
entropy of an infinitely massive, eternal Schwarzschild black hole by
(formally) evaluating the functional integral of Euclidean canonical
quantum gravity.  We find that the question of the finiteness of the
entropy can be answered only by understanding the renormalization of
the gravitational coupling $G$.

For completeness, we begin by formulating the functional integral
representation of the partition function.  If a spacetime manifold
${\cal M}$ is static, there exist coordinates $\{ x^{\mu} \}$ such
that the metric may be written
$$
g = g_{00}dx^0 \otimes dx^0 + g_{ij} dx^i \otimes dx^j \>,
\eqn\Lormet
$$
where $g_{\mu \nu}$ is independent of $x^0$, $i,j \in \{ 1,2,3 \}$,
and ${\cal M}$ has the topology $\IR \times \Sigma$.  Next, define
the Euclidean manifold $\Mbar$, with topology $S^1 \times \Sigma$, by
defining the Euclidean ``time'' coordinate $\theta = ix^0$ and
periodically identifying $\theta$ with period $\beta$.  The metric on
$\Mbar$ then has signature +4, and is written
$$
{\overline g} = -g_{00} d\theta \otimes d\theta + g_{ij} dx^i
\otimes dx^j \>.
\eqn\Riemet
$$
The partition function for fields $\phi$ propagating on ${\cal M}$
can be expressed as a functional integral
$$
Z(\beta) = {\cal N} \int {\cal D} [\phi] e^{-I [\phi]} \>,
\eqn\pfthree
$$
where ${\cal N}$ is a normalization factor and $I$ is the action of
the theory on the Euclidean manifold.  This method of calculation,
while derived from the Hamiltonian formulation of the quantum field
theory, has the advantage of being generally covariant.  The
formalism is extendable to include stationary as well as static
spacetimes.

For the theory of canonical quantum gravity, in which the metric is
one of the fields to be integrated over, the generalization of
equation \pfthree\ is taken to {\it define} the partition function
[\Gibhawk ].  The partition function is formally written as
$$
Z(\beta) = {\cal N} \int_{\cal F} {\cal D} [g] \int {\cal D}
[\phi] \exp (-I[g, \phi]) \>,
\eqn\pffour
$$
where the space ${\cal F}$ of Euclidean metrics is restricted by
boundary conditions, such as the total energy contained in spacetime
and behavior at infinity.  The Euclidean action functional $I$
appearing in the integral is
$$
I[g, \phi] = I_{EH} [g]+ I_{\phi} [\phi, g] \>,
\eqn\action
$$
where $I_{EH}$ is the Euclidean Einstein-Hilbert action
$$
I_{EH} [g] = {1 \over {16 \pi G_0}} \bigl ( - \int_{\Mbar} {\bf
\epsilon}_g R + 2\int_{\pa \Mbar} {\bf \epsilon}_h K \bigr ) \>,
\eqn\EinHil
$$
and $I_{\phi}$ is the action of the ``matter'' (non-gravitational)
fields $\phi$.  The bare gravitational coupling is explicitly denoted
by $G_0$.

The usual method of calculation of the partition function \pffour\ is
to find a manifold $\Mbar$ with metric ${\hat g}$ which is a
stationary point of the classical action and satisfies the boundary
conditions.  Then, writing an arbitrary metric $g$ as $g = {\hat g} +
f$, one quantizes the fluctuations $f$ and $\phi$ in the background
metric ${\hat g}$.  The action \EinHil\ can be expanded in powers of
$f$ as
$$
\eqalign{I[g, \phi] &= I[{\hat g} + f, \phi] \cr
&= I_{EH} [{\hat g}] + I_{\phi} [\phi, {\hat g}] + \int_{\Mbar}
{\bf \epsilon}_{\hat g} {{\delta I} \over {\delta g}} \bigr
\vert_{\hat g} f + \half \int_{\Mbar} {\bf \epsilon}_{\hat g}
{{\delta^2 I} \over {\delta g^2}} \bigr \vert_{\hat g} f^2
+ \ldots \>, \cr}
\eqn\acxpnd
$$
and the partition function \pffour\ can be written
$$
Z(\beta) = e^{-\beta F} = \exp (-I_{EH} [g]) Z'
\eqn\pffive
$$
where
$$
Z' = {\cal N} \int {\cal D} [f] \int {\cal D} [\phi] \exp \bigl
(-(I[{\hat g} + f,\phi] - I_{EH}[{\hat g}]) \bigr ) \>.
\eqn\Zdef
$$

To study the partition function for gravitational and matter fields
propagating outside an infinitely massive, eternal black hole, the
stationary point to expand around is a Euclidean continuation $\Rbar$
of Rindler space, with metric
$$
g = s^2 d\theta \otimes d\theta + ds \otimes ds + dx^2 \otimes
dx^2 + dx^3 \otimes dx^3 \>.
\eqn\Eucrinmet
$$
The Euclidean ``time'' coordinate $\theta$ is periodic with period
$\beta$, and we again restrict $x^2, x^3 \in \bigl [ -{L \over 2}, {L
\over 2} \bigr ]$ to regulate divergences due to the horizon area
$A$.  Now consider a subspace of constant $x^2$ and $x^3$.  This
subspace has the geometry of a cone with angle deficit $(2\pi - \beta
)$, and consequently has a conical singularity at the origin $s = 0$
proportional to $(2\pi - \beta )$.  However, $s = 0$ corresponds to
the event horizon of the black hole, which has no curvature
singularity in the Lorentzian geometry.  Imposing the condition that
this singularity be absent leads to the result that the correct
Euclidean continuation has $\beta = 2\pi$.  (This is one method of
deriving the Hawking-Unruh temperature, because even for a finite
mass black hole, the Rindler space approximation becomes arbitrarily
good as one gets close to the horizon.)  Indeed, only if $\beta =
2\pi$ will the metric be a stationary point of the Euclidean action
\EinHil .

Nevertheless, in order to obtain the entropy by means of equation
\entropytwo , one must know the partition function for $\beta$
slightly different from $2\pi$, so one must consider the geometries
containing conical singularities.  Indeed, it is shown in [\Claudio ]
that the entropy is conjugate to the deficit angle.  A Euclidean
continuation of Rindler space with period $\beta$ will be denoted
$\Rbar_{\beta}$.  It would seem that these geometries fail to be
stationary points of the functional integral.  Note, however, that
the derivative with respect to $\beta$ appearing in equation
\entropytwo\ is a partial derivative.  Thus, all other thermodynamic
variables must be held fixed, including the horizon area $A$.  The
condition that $A$ remain fixed must be implemented in the functional
integral.  This can be achieved by means of a Lagrange multiplier.
The effect of this multiplier is to insert an energy density along
the surface $s=0$, which is analogous to a cosmic string.  The
solution to the classical equations of motion will therefore have a
conical singularity at $s=0$.  Thus the Euclidean continuations for
$\beta \neq 2\pi$ are stationary points of the Euclidean functional
integral subject to the constraint of constant horizon area.

Consider now the factor $\exp (-I_{EH} [{\hat g}])$ in equation
\pffive , which gives the contribution to the entropy from the
classical geometry.  The effect of the curvature singularity is that
$$
\int_{\Rbar_{\beta}} {\bf \epsilon}_{\hat g} R = 2 A (2\pi -
\beta ) \>.
\eqn\Rcond
$$
Thus the Einstein-Hilbert action is
$$
S_{EH}[{\hat g}] = -{{(2\pi - \beta ) A} \over {8 \pi G_0}} =
\beta F \>.
\eqn\SEH
$$
{}From equation \SEH\ one obtains the Bekenstein-Hawking formula for
the entropy per unit area,
$$
{\sigma \over A} = {1 \over {4G_0}} \>.
\eqn\Rent
$$
Note, however, that this entropy does not have any obvious origin in
the counting of quantum states, since it arises simply from the
classical action of the conical Rindler background.

To proceed, we must (formally) calculate the functional integral $Z'$
in equation \pffive .  The actual quantization of the theory defined
by equation \pffive\ is a tricky subject, but we shall find that we
do not need more than the most basic of results.  First note from
equation \acxpnd\ that, neglecting the fluctuations $f$, $Z'$ reduces
to the partition function for matter fields propagating in the
Rindler background.  For the case of a single free scalar field, this
was calculated in Chapter 2 .  From the point of view of the
Euclidean functional integral, the computation in Chapter 2 is
equivalent to considering the free matter field on a fixed Rindler
background $\Rbar_{\beta}$ with conical deficit $2\pi - \beta$.  The
functional integral can be represented in terms of first quantized
particle paths, according to a standard prescription.  In higher
orders of perturbation theory, the paths branch to form Feynman
diagrams.  However, the only diagrams which can contribute to the
entropy are those which intersect or encircle the horizon at $s=0$.
To see why, consider first the lowest order diagrams, which are
simple closed curves.  A single loop with fixed ``center of mass''
which does not intersect or encircle $s=0$ is insensitive to the
deficit angle.  Summing over all such loops leads to an integration
of the center of mass of each loop over all angular positions, and
the result is proportional to $\beta$.  Accordingly, it will give no
contribution to the entropy.  A loop which encircles the singularity
represents a real particle propagating outside the black hole, and
this state certainly contributes positively to the entropy.  The
divergence in equation \phient\ is evidently due to very small loops
which encircle $s=0$.  It is for this reason that the entropy is
localized near the horizon.  The point is that calculating the
partition function by evaluating the Euclidean functional integral on
a conical background provides a unified way of obtaining the entropy
due to both the classical geometry and the quantum corrections.
However, thus far only the quantum corrections have a clear
interpretation in terms of the counting of states.

Integrating out all the matter fields and tree level gravitons gives
a contribution proportional to ${1 \over \varepsilon^2}$.  Including
this term, the entropy per unit area is
$$
{\sigma \over A} = {1 \over 4} \bigl ( {1 \over G_0} + {C \over
{90 \pi \varepsilon^2}} \bigr ) \>,
\eqn\entapprox
$$
where $C$ is a constant which depends on the matter content of the
theory.  As we shall see in what follows, the same quadratic
divergence enters into the renormalization of the gravitational
coupling in the effective action.

We now proceed to prove this statement.  After integrating out the
matter fields and fluctuations of the metric, $Z'$ has the form $Z' =
\exp (-W')$, where on general grounds $W'$ must be a
diffeomorphism-invariant functional of the background metric
$g$.\foot{Here and henceforth we drop the caret over the background
metric $g$.}  $W'$ will contain all possible covariant terms, and may
be expanded in powers of the Riemann tensor and its derivatives as
$$
W'[g] = \int_{\Mbar} {\bf \epsilon}_g \bigl [ -{1 \over {16\pi}}
a R + Q(R) \bigr ] \>.
\eqn\Wexp
$$
Here $a$ is a constant and $Q$ contains all other induced covariant
terms.  We neglect a possible renormalization of the cosmological
constant.  The effect of $a$ is to renormalize the value of the
gravitational coupling in the effective action from $G_0$ to $G_R$,
given by
$$
{1 \over G_R} = {1 \over G_0} + a \>.
\eqn\Gren
$$

The next step is to evaluate equation \Wexp\ for Euclidean Rindler
space $\Rbar$.  To regulate the curvature singularity at the origin,
define $R = (2\pi - \beta ) f$, where $f$ is a smooth function
supported only on an $\varepsilon-$neighborhood of the origin.  The
condition that \Rcond\ be satisfied means that $f$ must satisfy
$$
\int_{\Rbar_{\beta}} {\bf \epsilon}_g f = 2A \>.
\eqn\fcond
$$
We also require that the scale of variation of $f$ is independent of
the conical angle, so that derivatives of $f$ do not introduce
additional dependence on $\beta$.  Now consider the possible types of
terms that can appear in $Q$.

1.  Any local or nonlocal term with $n \geq 2$ powers of the Riemann
tensor $R_{\alpha \beta \mu \nu}$ will be proportional to $(2\pi -
\beta )^n$.  This includes terms with arbitrary numbers of derivative
operators acting on $R_{\alpha \beta \mu \nu}$.  Their contribution
to $W'$ may be represented as
$$
- \sum_{n=2}^{\infty} b_n (2\pi - \beta )^n \>,
\eqn\bfirst
$$
where the $b_n$ are constants.

2.  Now consider terms linear in $R_{\alpha \beta \mu \nu}$, with
arbitrary derivatives acting on them.  For example, consider
$$
I = \int_{\Rbar} {\bf \epsilon}_g (\nabla^2) R \>.
\eqn\proofone
$$
Since $R$ is now a smooth function, by use of Stokes' theorem
equation \proofone\ can be rewritten as an integral over the boundary
of $\Rbar$,
$$
I = \int_{\pa {\Rbar}} {\bf \epsilon}_h n^{\mu} \nabla_{\mu} R \>,
\eqn\prooftwo
$$
where $n$ is a unit vector normal to $\pa \Rbar$.  But $R$ vanishes
outside a small neighborhood of the origin, so the integral $I$
vanishes.  It is obvious that all such terms will vanish after
integration by parts.

Due to the rapid falloff of the Green functions in four dimensions,
nonlocal terms proportional to one power of $R_{\mu \nu \alpha
\beta}$ will not appear, and the above list covers all possible
terms.  Thus, using the condition \Rcond , the full Helmholtz free
energy $\beta F = S_{EH}[g] + W'[g]$ can be written
$$
\beta F = -\sum_{n=1}^{\infty} b_n (2\pi - \beta )^n \>,
\eqn\freen
$$
where $b_1 = {A \over {8\pi G_R}}$ comes only from the
Einstein-Hilbert term.  The entropy is therefore
$$
\sigma(\beta) = {{\beta A} \over {8\pi G_R}} + \sum_{n=1}^{\infty}
(b_n + \beta (n+1) b_{n+1}) (2\pi - \beta )^n \>.
\eqn\entone
$$
Setting $\beta = 2\pi$, equation \entone\ reduces to the
Bekenstein-Hawking entropy \BHent , but with the renormalized
gravitational coupling $G_R$ given by equation \Gren .

Thus we arrive at the conclusion that for the case of canonical
quantum gravity coupled to matter fields, {\it the expression \BHent\
for the Bekenstein-Hawking entropy of the fields propagating outside
a black hole is a general result, but the gravitational coupling
appearing in equation \BHent\ is the renormalized gravitational
coupling $G_R$ given by equation \Gren .}  Comparing equations \Gren\
and \entapprox , we see that the divergences in the entropy are the
same divergences which renormalize the gravitational coupling.  In
particular, this means that the question of the finiteness of the
entropy of the black hole is inextricably intertwined with the
renormalization of the theory.  Canonical quantum gravity is
nonrenormalizable, and it is often the case that the only consistent
quantum field theory that can be obtained from a nonrenormalizable
theory is a free field theory.  If this is the case with canonical
quantum gravity, then $G_R = 0$, and the entropy diverges.

%
%
\chapter{Two dimensional models}

Because of the large amount of attention that has recently been
focused on two dimensional toy models of black holes, it is of
interest to examine how quantum corrections affect the entropy of a
two dimensional black hole.  In the following it will be shown that
in the two dimensional model proposed by Callan, Giddings, Harvey,
and Strominger (CGHS) [\CGHS ], the divergence in the entropy of
scalar fields moving in a black hole background is {\it not} the same
as the divergence which renormalizes the gravitational coupling.
Instead, it provides an infinite zero point entropy, which
corresponds to an infinitely degenerate ground state and probably a
theory which loses information.

The CGHS model is defined by the action functional
$$
S_{CGHS} = {1 \over {2\pi}} \int \epsilon_g \bigl ( e^{-2\phi}
[R + 4(\nabla \phi)^2 + 4\lambda^2 ] - {1 \over 2} (\nabla f)^2
\bigr ) \>,
\eqn\Scghs
$$
where $g$ , $\phi$, and $f$ are the metric, dilaton, and matter
fields, respectively, and $\lambda^2$ is a cosmological constant
which defines a length scale for the theory.  The classical theory
defined by the action \Scghs\ has eternal black hole solutions.
Defining light cone coordinates $x^{\pm}$ and choosing the line
element to have the form $ds^2 = - e^{2\rho} dx^+ dx^-$, these
solutions are given by
$$
e^{-2\rho} = e^{-2\phi} = {M \over \lambda} - \lambda^2
x^+ x^- \>,
\eqn\met
$$
where $M$ is the black hole mass.  The future horizon is the curve
$x^- = 0$.

The CGHS theory can be viewed as an effective action for radial modes
of near extreme, magnetically charged black holes in four dimensional
dilaton gravity [\CGHS].  Using this correspondence, the area of the
black hole is defined to be $\lambda^{-2} e^{-2\phi}$ evaluated at
the horizon, or $A = M \lambda^{-3}$.

As for any non extreme black hole, the behavior of the geometry near
the horizon is closely approximated by Rindler space.  Taking the
limit $M \rightarrow \infty$, a calculation of the entropy of the
scalar field $f$ may be performed using techniques analogous to those
used in Chapter 2.  Due to the bad infrared behavior of scalar fields
in two dimensions, in addition to the horizon cutoff $\varepsilon$
one must also introduce an infrared cutoff $\ell$.  The entropy is
found to be
$$
\sigma = {1 \over 6} \log \bigl ( {\ell \over \varepsilon}
\bigr ) \>.
\eqn\fent
$$
Note that this entropy is {\it not} proportional to the horizon area.
 Instead, it represents an infinite additive constant to the entropy.
 Indeed, it can be seen that the entropy of the $f$ field cannot be
proportional to the area because of the way $f$ couples to the
dilaton.  From the point of view of the four dimensional theory, the
result \fent\ occurs because the truncation of all but the
spherically symmetric modes is a violent reduction of the number of
degrees of freedom of the theory.

The origin of this entropy can also be understood by examining the
effective action obtained after integrating out the $f$ field.  This
action is given by the original action \Scghs\ plus a Liouville
action, which can be written using the above metric as
$$
\eqalign{S_L &= -{1 \over {12\pi}} \int d^2 x \bigl ( \rho - \log
(\ell / \varepsilon) \bigr ) (\nabla)^2 ( \rho - \log (\ell /
\varepsilon) \bigr ) \cr
&= - {1 \over 96\pi} \int \epsilon_g R {1 \over {\nabla^2}} R -
{{\log (\ell / \varepsilon )} \over {12\pi}} \int \epsilon_g R
\>. \cr}
\eqn\louie
$$
Note that the $\rho$ field only appears in the combination $\rho -
\log (\ell / \varepsilon)$.  The first term in equation \louie\ is
the familiar correction to the classical action, and is responsible
for the Hawking radiation from the two dimensional black hole.  The
second term is proportional to the Euler class, and is the term which
gives rise to the divergent entropy of the scalar field.  Indeed,
being careful with factors of $\pi$ which enter into definitions of
energy in the theory, the entropy \fent\ can be read off from the
second term in equation \louie .  It should be pointed out that the
first term in the action \Scghs\ gives rise to the classical
contribution to the entropy, and as with the four dimensional black
hole, there is no obvious counting of states associated with this
entropy.

Since the integrated curvature is a topological invariant, the second
term in the effective action \louie\ plays no part in the dynamics of
the theory, and is usually ignored.  Moreover, its contribution to
the entropy is as an additive constant, which has no thermodynamic
significance.  However, the presence of this term does have
information theoretic significance.  Due to this infinite additive
constant, there is no mechanism to prevent the black hole from
accumulating an arbitrarily large amount of information near the
horizon.  It is therefore highly plausible that information is, in
fact, lost in two dimensional theories, or that black hole remnants
exist in these theories, which amounts to the same thing.
Nevertheless, we have seen that a truncation to only spherically
symmetric modes does great violence to a four dimensional theory -
the entire renormalization structure is profoundly changed.  For this
reason, we conclude that two dimensional theories do not possess
enough degrees of freedom to be viable models of four dimensional
gravity.

%
%

\chapter{Superstrings and black hole entropy}

Having learned in Chapter 3 that the question of black hole entropy
is related to the ultraviolet divergence structure of our theory of
quantum gravity, the next step is to look for theories for which this
divergence structure can be understood.  A natural candidate for
examination is the theory of superstrings propagating in a background
spacetime.

There are two essential points we will establish in the remainder of
this paper.  The first point is that in string theory, unlike
canonical quantum gravity, the entire entropy per unit area of a
horizon can be attributed to identifiable quantum states.  In other
words, superstring theory is an ``induced'' theory of the
gravitational effective action, in which all the terms in the action
arise from integrations over fluctuations in the presence of a
background geometry.  This is true even for the ``classical'' or tree
graph action, which is generated by integration over genus zero
surfaces.  Computing these fluctuations in a conical background
requires a definition of the string theory ``off-shell''.  Such off
shell continuations introduce ambiguities and divergences into the
world sheet sigma model.  It will be shown, however, that the entropy
is entirely independent of these ambiguities.

The second important point to establish is that string theory is also
ultraviolet finite, and therefore leads to a finite entropy per unit
area.  This follows from the fact that the renormalization of the
string coupling constant $\kappa$ (in units of the string tension) is
finite order by order in perturbation theory.

In the present state of development of string theory, it is not
possible to begin with a Hamiltonian, solve for the eigenvalues, and
compute the partition function using equation \pfone .  Therefore, we
must make an ansatz that the definition \pfthree\ of the partition
function holds in string theory, by which we mean that the logarithm
of the partition function is given in terms of a sum of string
Feynman graphs in the appropriate Euclidean continuation of the
spacetime manifold.  This ansatz is supported by results in
[\Polchinski ].

Using this ansatz, we will show how superstring theory resolves the
puzzles behind the Bekenstein-Hawking entropy.  We show that the
contribution to the entropy per unit area from genus zero string
graphs is the leading order ``classical'' term in the
Bekenstein-Hawking formula.  This classical term thus has a manifest
origin in the counting of quantum states.  Next we find that the
Bekenstein-Hawking result holds to all orders in superstring
perturbation theory, and that the renormalized gravitational coupling
appearing in the formula is {\it finite.}

The starting point for our discussion is the two-dimensional
supersymmetric sigma model describing the propagation of superstrings
in a background spacetime metric $g$.  The generating functional for
the two-dimensional superconformal field theory on a world sheet of
genus $n$ is
$$
Z^{(n)} = {{\kappa_0^{2(n-1)}} \over {\Vol ({\cal G})} } \int
{\cal D} [e] \int {\cal D} [\chi] \int {\cal D} [X] \int {\cal D}
[\Psi] \exp (-I[X, \Psi ;e, \chi]) \>,
\eqn\gfone
$$
where $X^{\mu}$ and $\Psi$ are the bosonic and fermionic coordinates
of the superstring, respectively, $e$ is the world sheet zweibein,
and $\chi$ is the gravitino.  $\kappa_0$ is the bare string coupling,
and ${\cal G}$ denotes the symmetry group of the two dimensional
action $I$, which includes diffeomorphisms, superconformal
transformations, and an on shell local supersymmetry.  The first step
is to gauge fix the world sheet zweibein to $e = e^{\Lambda} {\hat
e}$ and the gravitino to $\chi = \rho \lambda$, where ${\hat e}$ is a
fiducial zweibein, $\rho$ are the two dimensional Dirac matrices, and
$\lambda$ is a Grassmann variable.  This introduces reparametrization
ghosts $b$, $c$, $\beta$, and $\gamma$, and equation \gfone\ becomes
$$
Z^{(n)} = \kappa_0^{2(n-1)} {{\int {\cal D} [\Lambda, \lambda]}
\over {\Vol (\SC )}} \int_{F_n} d^{2m_n} \tau {{\int {\cal D} [X]
{\cal D} [\Psi] {\cal D} [b, c, \beta, \gamma]} \over {\Vol
(\Omega )}} \exp (-I[X,b,c, \beta, \gamma ; e^{\Lambda} {\hat e},
\rho \lambda]) \>.
\eqn\gftwo
$$
Here SC denotes the group of superconformal transformations, $F_n$ is
a fundamental region for the integration over the $2m_n$ supermoduli
$\tau$, and $\Omega$ denotes the additional subgroup of symmetries
which remains after the gauge fixing.  $\Omega$ is generated by the
conformal Killing vectors and spinors.  We imagine regulating the two
dimensional field theory by replacing the world sheet by a finite
lattice.  The volume of the group $\Omega$ is then also naturally
regulated [\Junone, \Tseyone ].

In order to cancel the integrals over $\Lambda$ and $\lambda$ against
Vol(SC), the theory must actually be superconformally invariant.  For
strings propagating in flat $D$ dimensional Euclidean space with no
background fields, the condition for superconformal invariance is
that the dimension take the value $D=10$.  For a general
supersymmetric sigma model, the conditions for superconformal
invariance are more complicated, and in addition to the condition
$D=10$, the spacetime fields must satisfy the equations of motion of
a spacetime action.  The superconformal invariance conditions imply
the vanishing of the beta functions for the two dimensional quantum
field theory, so the theory is ultraviolet finite.

To study the statistical mechanics of superstrings near massive,
eternal black holes, we shall be interested in calculating the
$Z^{(n)}$ for superstrings propagating in Euclidean Rindler space.
In order to calculate the entropy, however, a small conical
singularity must be introduced, and the resulting space violates the
conditions for superconformal invariance of the theory.  Therefore, a
prescription must be followed for calculating the off shell
generating functionals, and we will continue to use \gftwo\ as our
definition of $Z^{(n)}$.  It will be found, however, that the entropy
is independent of the prescription used.

Consider first the case of genus zero.  After integrating over the
world sheet fields and dividing out the volume of $\Omega$, equation
\gftwo\ for $Z^{(0)}$ takes the form
$$
Z^{(0)} = \kappa_0^{-2} {{\int {\cal D} [\Lambda, \lambda]} \over
{\Vol (\SC )}} F(g; \varepsilon, \Lambda, \lambda) \>,
\eqn\Znot
$$
where $F$ is a generally covariant functional of the background
metric $g$, and also depends on the world sheet regulator parameter
$\varepsilon$ and the superconformal parameters $\Lambda$ and
$\lambda$.  The basic structure of $F$ can be determined by quite
general arguments.  To begin with, for a fixed value of the
regulator, $F$ may be expanded as a sum of integrals of powers of the
Riemann tensor and its derivatives.  To see why, we return to our
regulation of the string theory in which the world sheet is replaced
by a finite lattice of points.  Then the bosonic part of the
functional integral becomes a product of ordinary coupled Gaussian
integrals.  Because of the exponential damping, the integrand tends
to zero quickly when the integration variables are distantly
separated.  In particular, the image of the worldsheet in the target
spacetime will have an extent of order $\ell^2 \sim \log({1 \over
\varepsilon})$.  Thus, for a fixed value of the regulator, there is
no way to introduce any nonlocal behavior with extent larger than
${\cal O}(\ell)$ into the generating functional, and a series
expansion of the type described above will be possible.  It should
also be pointed out that the quantity $Z^{(0)}$ is not the effective
action for string theory as defined by Tseytlin in [\Tseyone ], but
is simply related to it, as is demonstrated in Appendix A.

The coefficients of the terms in the expansion of $F$ will depend on
$\varepsilon$, $\Lambda$, and $\lambda$, and will in general diverge
as $\varepsilon$ goes to zero.  This is one of the difficulties
involved in defining string theory off shell.  It is shown in
Appendix A, however, that the coefficient of the term $\int
\epsilon_g R$ is independent of $\varepsilon$, $\Lambda$, and
$\lambda$.  For this term the integral over the superconformal
parameters simply cancels Vol(SC), and so $Z^{(0)}$ can be written
$$
Z^{(0)} = - \kappa_0^{-2} \int \epsilon_g R + \kappa_0^{-2} {{\int
{\cal D} [\Lambda, \lambda]} \over {\Vol (\SC )}} Q(g; \varepsilon,
\Lambda, \lambda) \>,
\eqn\Znotagain
$$
where $Q$ contains all the other terms in $F$.

Although it is apparent that a unique definition of the off shell
amplitude does not exist, it is obvious that the first term in
equation \Znotagain\ governs the low energy scattering of gravitons,
and that its coefficient can be related in the usual way to the bare
gravitational coupling.

The genus zero generating functional $Z^{(0)}$ has been written down
for a ten dimensional background metric, but we want to study four
dimensional physics, so we must introduce a compactification scheme.
For simplicity, we will consider a target space $\Tbar$ which is a
product manifold $\Mbar \times K$, where $\Mbar$ is a four
dimensional manifold coordinatized by $\{ x^i \}_{i=1}^4$ (which will
eventually be identified with four dimensional Euclidean Rindler
space), and $K$ is a $D-4$ dimensional compact manifold coordinatized
by $\{ x^i \}_{i=5}^D$ and having no intrinsic curvature.  The metric
on $\Tbar$ is block diagonal, decomposing into a metric $g^{(4)}$ on
$\Mbar$ and a metric $g^{(D-4)}$ on $K$.  A simple choice for $K$ is
the product manifold $(S^1)^{D-4}$, with $x^i \in [0, L_i ]$ for $i
\in \{5, \ldots , D \}$, and $g_{ij}^{(D-4)} = \delta_{ij}$.  The
$L_i$ are taken to be on the order of $\sqrt{\alpha '}$.  The genus
zero generating functional can now be written
$$
Z^{(0)} = - {1 \over {16\pi G_0}} \Bigl ( \int_{\Mbar}
\epsilon_{g^{(4)}} R + {{\int {\cal D} [\Lambda, \lambda]} \over
{\Vol (\SC )}} \, \int_{\Mbar} \epsilon_{g^{(4)}} Q(g^{(4)};
\varepsilon, \Lambda, \lambda) \Bigr ) \>,
\eqn\Znottwo
$$
where the bare four dimensional gravitational coupling is
$$
G_0 = {\kappa_0^2 \over {16\pi \int_K \epsilon_{g^{(D-4)}}}} \>.
\eqn\gravcoup
$$

All that is left now is to specify the four dimensional manifold
$\Mbar$ and the background metric.  For our purposes, the manifold
$\Mbar$ is taken to be Euclidean Rindler space $\Rbar_{\beta}$, which
has an angle deficit $(2\pi - \beta)$, and metric $g^{(4)}$ given by
equation \Eucrinmet .  From here the argument proceeds exactly as in
section 3, where we found that the entropy per unit area at $\beta =
2\pi$ depended only on the coefficient of the integral of $R$, and we
obtain the result that the entropy per unit area obtained from genus
zero string graphs is given by the Bekenstein-Hawking formula,
$$
{\sigma \over A} = {1 \over {4G_0}} \>.
\eqn\zgent
$$
It should by now be apparent that this result does not depend on the
exact definition of off shell superstring generating functionals,
because changes in the prescription for off shell functionals can
only influence the result through the terms which depend on the
regulator or the superconformal parameters.  These terms all give
contributions to the entropy which vanish when one sets $\beta =
2\pi$.

Assuming that superstring theory can be written in a Hamiltonian
formulation, it is surprising that the entropy we have obtained is
well defined only for the Rindler temperature $\tau_{R} = {1 \over {2
\pi}}$.  For ordinary systems, one can calculate $\Tr ( \EXP ( -\beta
H ) )$ for any value of $\beta$.  A similar situation occurs,
however, when one tries to calculate the thermodynamics of a
sufficiently large gravitating system.  Here one is foiled by the
fact that the long range gravitational field leads to the Jeans
instability, and the thermal ensemble is not well defined.  To see
the connection with the above calculation, notice that the world
sheet ultraviolet regulator $\varepsilon$ acts as a spacetime
infrared regulator.  For example, a string graph will have an extent
of order $\sqrt{\log ( {1 \over \varepsilon})}$, as mentioned
previously.  The above results suggest that an infrared instability
also occurs in our calculation, and that the regulator $\varepsilon$
controls it, allowing one to perform statistical mechanical
calculations at an arbitrary temperature.  The thermodynamics of the
system, and in particular the entropy, will therefore depend on this
infrared regulator.  The entropy can be made well defined by taking
away the regulator only for $\beta = 2\pi$.

%
%
\chapter{String states on the horizon and the finiteness of the
entropy}

The important thing to note about the result \zgent\ is that in
superstring theory the leading order, ``classical'' contribution to
the Bekenstein-Hawking entropy arises explicitly from an integration
over string configurations, namely, those described by genus zero
string graphs.  It is precisely because of its origin in the counting
of string states that the entropy is independent of the exact off
shell prescription used - the entropy is an ``on shell quantity''.
The nature of the states contributing to the entropy \zgent\ will now
be determined.

The bosonic part of a general genus zero superstring graph is a
continuous map from $S^2$ into the target space $\Rbar_{\beta} \times
K$, as shown in Figures 1 and 2.  Just as in the case of the first
quantized particle paths described in Chapter 3, not all such graphs
can contribute to the entropy.  Only the graphs which intersect the
conical singularity at $s = 0$ can contribute.  This can be
understood as follows.  Consider deleting the subspace $s = 0$, so
that the resulting Euclidean manifold has topology $S^1 \times \IR^3
\times K$.  Since $S^2$ is simply connected and the mapping is
continuous, the action of a graph cannot depend on $\beta$.  Summing
over all string graphs involves an integration over the angular
location of a given graph, and thus will be proportional to $\beta$.
This $\beta$ dependence will be cancelled when one obtains the
Helmholtz free energy from the generating functional.  On
differentiating $F$ with respect to $\beta$, the sum vanishes.
%
%
\vskip 15pt
\vbox{{\centerline{\epsfsize=3.0in \hskip 2cm \epsfbox{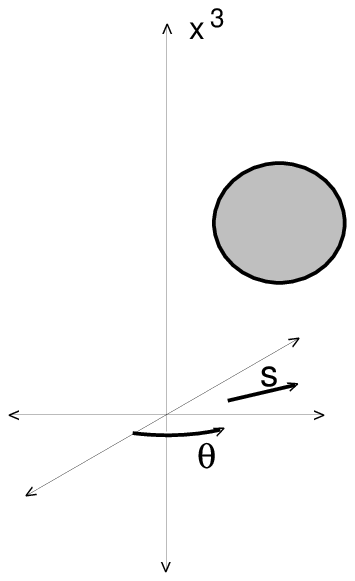}}}
\vskip 12pt
{\centerline{\tenrm FIGURE 1. A genus zero string graph which does
not contribute to the entropy.}}
{\centerline{\tenrm All other directions besides $s$, $\theta$, and
$x^3$ are suppressed.}}
\vskip 15pt}
\vbox{{\centerline{\epsfsize=3.0in \hskip 2cm \epsfbox{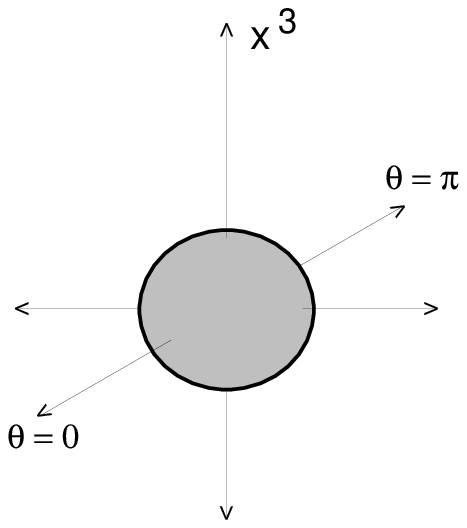}}}
\vskip 12pt
{\centerline{\tenrm FIGURE 2. A genus zero string graph which does
contribute to the entropy.}}
\vskip 15pt}

An example of a genus zero graph which does contribute to the entropy
is given in Figure 2.  To understand what state this corresponds to,
we consider a slice of constant Euclidean time, shown in Figure 3.
It is apparent that the states we are counting are closed
superstrings which lie partially behind the horizon.  An observer
outside the black hole describes these states as open superstrings
with both ends frozen to the horizon, but which are free to interact
with each other and with superstrings propagating outside the black
hole.  Thus we have proved our first point, that the classical
contribution to the Bekenstein-Hawking entropy arises from counting
identifiable states of superstrings frozen to the horizon.
%
%
\vskip 15pt
\vbox{{\centerline{\epsfsize=3.0in \hskip 2cm \epsfbox{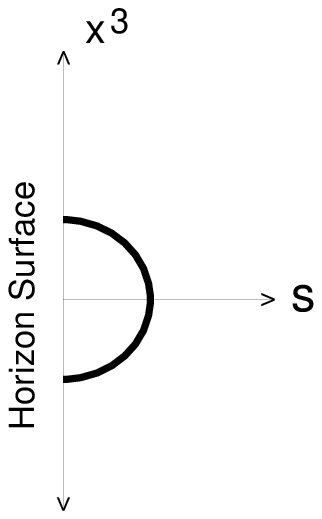}}}
\vskip 12pt
{\centerline{\tenrm FIGURE 3. A slice of constant Euclidean time
showing an open string}}
{\centerline{\tenrm with both ends attached to the horizon.}}
\vskip 15pt}

Now consider contributions from higher genus graphs.  After
performing the functional integrals over the superstring coordinates
and the ghosts, each $Z^{(n)}$ can be expressed in a form similar to
equation \Znottwo .  Using the now familiar arguments of Chapters 3
and 5, the only terms we are interested in are the integrals of the
scalar curvature, which are superconformally invariant, so we drop
the rest.  Denote the genus $n$ contribution as
$$
J^{(n)} = - a_n \int \epsilon_{g^{(4)}} R \>.
\eqn\Jn
$$
The effect of the coefficients $a_n$ is to renormalize the bare
gravitational coupling to its renormalized value $G_R$.  It is well
known that the integrals over the world sheet moduli are the
analogues of the integrals over Feynman-Schwinger parameters in
quantum field theory.  In field theory, these integrals lead to
divergences, such as those which lead to infinities in the entropy.
It is also well known that in superstring theory the dangerous
regions of moduli space are eliminated, and the coefficients $a_n$
are finite [\GSWtwo ].  Thus the renormalized coupling constant is
finite.  We therefore arrive at our second conclusion, that the
entropy per unit area of a horizon is {\it finite\/} to all orders in
superstring perturbation theory, and is given  by
$$
{\sigma \over A} = {1 \over {4G_R}} \>,
\eqn\fdsustrenttwo
$$
where $G_R$ is the finite renormalized value of the four dimensional
gravitational coupling.

As an aside, for certain superstring theories the renormalization of
the gravitational coupling is in fact zero.  This is the case when,
after compactifying the theory down to four dimensions, the theory
exhibits an $N=4$ supersymmetry.  If the supersymmetry is broken to
some lower $N$, then the renormalization of the gravitational
coupling is finite but nonzero [\Vadim ].

As with the genus zero case, only certain higher genus graphs
contribute to the entropy.  By an argument similar to the above, one
can easily see that the only graphs which contribute to the entropy
must encircle or intersect the conical singularity at $s=0$.
Examples of such genus one graphs are given in Figures 4 and 5.
Figure 4 depicts a closed superstring which remains outside the black
hole for all time.  Such graphs represent states in which the entire
string remains outside the horizon.  Figure 5 describes a process in
which an open string emits and reabsorbs a closed string.

Now an interesting puzzle arises.  Consider the fact that in $N=4$,
$D=4$ compactifications of superstring theory (as well as ordinary
$N=4$, $D=4$ supergravity theories) the higher loop corrections to
the gravitational coupling vanish.  By our result in section 5
(section 3), this implies that the corrections to the entropy vanish
as well.  The fact that the gravitational coupling is not
renormalized can be attributed to a cancellation between fermionic
and bosonic degrees of freedom in Feynman graphs.  However, if one
performs an ordinary statistical mechanical calculation of the
entropy that counts physical states, then these states, whether
bosonic or fermionic, always contribute positively to the entropy.
The question is, from the point of view of the statistical mechanical
calculation, where do the necessary negative contributions to the
entropy come from?
%
%
\vskip 15pt
\vbox{{\centerline{\epsfsize=3.0in \hskip 2cm \epsfbox{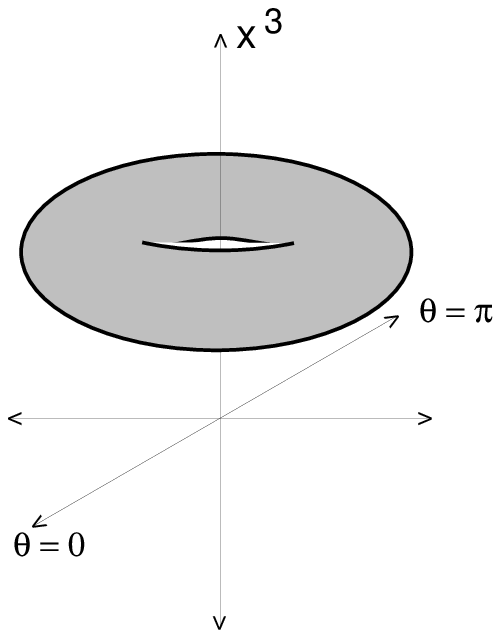}}}
\vskip 12pt
{\centerline{\tenrm FIGURE 4. A genus one string graph which
contributes positively to the entropy.}}
\vskip 15pt}
\vbox{{\centerline{\epsfsize=3.0in \hskip 2cm \epsfbox{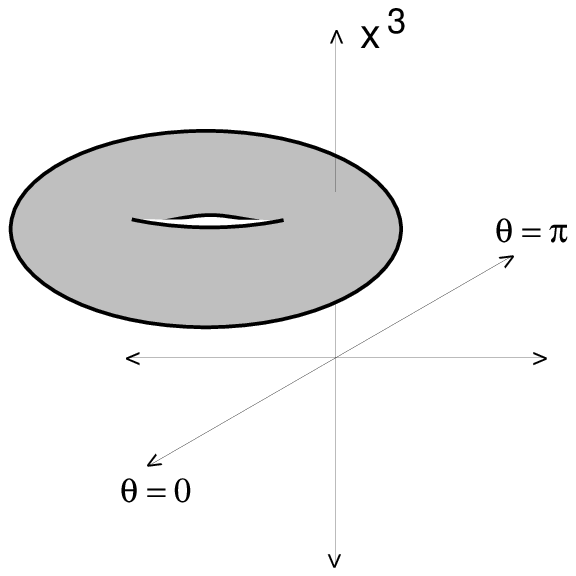}}}
\vskip 12pt
{\centerline{\tenrm FIGURE 5. A genus one string graph describing a
coupling between strings on the horizon}}
{\centerline{\tenrm  and strings outside the horizon.}}
\vskip 15pt}

In superstring theory, the resolution of this puzzle comes from genus
one graphs which intersect the horizon as in Figure 5.  These graphs
do not represent the contribution of either open strings attached to
the horizon, or closed strings outside the horizon.  They describe
processes which couple the open and closed string sectors.  In other
words, these graphs describe an interaction between the degrees of
freedom outside the horizon and those that make up the horizon.
These graphs do not have to contribute positively to the entropy, and
in fact must provide the negative contributions to the entropy which
cancel the positive contributions from the closed string graphs shown
in Figure 4.

%
%
\chapter{Conclusions}

Let us now take stock of what has been learned.  In Chapter 2 it was
shown that quantum fields propagating in a fixed four dimensional
Rindler space background have an entropy per unit area which diverges
near the horizon.  This divergence is due to an infinitely large
number of states having arbitrarily small energy.  In Chapter 3 it
was shown that this divergence is the same divergence which
renormalizes the gravitational coupling in the effective action of
canonical quantum gravity.  For this reason, the divergence in the
entropy cannot be understood without an understanding of the
ultraviolet behavior of the theory.  In Chapter 4, two dimensional
toy models of black holes were examined, and it was found that the
matter fields in the theory contribute an infinite additive constant
to the entropy.  It is therefore reasonable to expect that these
models do in fact exhibit information loss.  It was argued, however,
that these models do not possess enough degrees of freedom to be good
models of four dimensional black holes.

In the last two chapters we showed how superstring theory, in its
perturbative formulation, resolves the puzzles of the
Bekenstein-Hawking entropy.  The leading order classical contribution
to the entropy per unit area was shown to arise from superstrings
which lie partially behind the horizon, and which act like open
strings with both ends attached to the horizon.  This result gives a
new physical picture of a black hole - that of a surface covered with
bits of string which are free to interact with each other and strings
outside the black hole.  These bits of string give rise to a
microstructure on the horizon, which can be thought of in the field
theory limit as an additional set of degrees of freedom, which one
could call a stretched horizon [\STU ].

In addition, the entropy per unit area has been shown to be finite to
all orders in superstring perturbation theory, and to be given by the
Bekenstein-Hawking formula, equation \fdsustrenttwo .  This is
because the renormalized gravitational coupling is finite.
Therefore, in superstring theory, a black hole cannot absorb an
infinite amount of information, and must emit the information in the
form of Hawking radiation.

The results obtained in this paper only apply in the limit of
infinite black hole mass.  For large but finite black hole mass $M$,
one would in general expect corrections to the black hole entropy
which are of ${\cal O}(1/M)$ compared to the Bekenstein-Hawking
formula.  We do not doubt the existence of these terms, and it is our
belief that these terms will be finite in superstring theory.
Nevertheless, given a sufficiently large black hole, Hawking's
information problem can be formulated without mention of these terms,
so the resolution of the question of information loss should depend
only on the finiteness of the Bekenstein-Hawking term.

It should be emphasized that the validity of our result rests on the
ansatz for the logarithm of the partition function used in Chapter 5.
 A true calculation of the entropy would require an actual
enumeration of states, for which a Hamiltonian is needed.  In the
absence of a Hamiltonian, however, the calculation above appears to
be the most viable alternative.  If superstring theory is truly
consistent, it is reasonable to expect that the Hamiltonian
calculation will reproduce the results obtained here.

%
%

\appendix

In this appendix we show that the coefficient of the term $\int
\epsilon_g R$ in $F$ is independent of $\varepsilon$, $\Lambda$, and
$\lambda$.  We will use results obtained by Tseytlin in reference
[\Tseyone ].  Define the quantity ${\tilde Z}^{(0)}$ using equation
\gftwo\ by
$$
Z^{(0)} = \kappa_0^{-2} {{\int {\cal D} [\Lambda, \lambda]} \over
{\Vol (\SC )}} {1 \over {\Vol (\Omega )}} {\tilde Z}^{(0)} \>.
\eqn\appone
$$
so that ${\tilde Z}^{(0)}$ is the regulated generating functional for
the two dimensional field theory, computed in a particular conformal
gauge, and without the volume of $\Omega$ removed from it.  It is
shown in [\Tseyone ] that $\Vol (\Omega ) \propto \log (\varepsilon
)$, and that there exists a field redefinition such that
$$
{{\pa {\tilde Z}^{(0)}} \over {\pa (\log (\varepsilon ))}} = I_{eff}
\>,
\eqn\apptwo
$$
where $I_{eff}$ is the spacetime action which generates equations of
motion equivalent to the superconformal invariance conditions.
$I_{eff}$ has an expansion of the form
$$
I_{eff} =  c \int \epsilon_g \Bigl ( -R + {4 \over {(D-2)}} (\nabla
\Phi)^2 +
{1 \over 3} \exp \bigl ( {{8\Phi} \over {D-2}} \bigr ) H^2 + \alpha '
Q(g; \alpha ', \varepsilon, \Lambda, \lambda) \Bigr ) \>,
\eqn\appthree
$$
where $c$ is a constant and $Q$ contains terms which are higher order
in $R_{\alpha \beta \mu \nu}$, with coefficients that depend on
$\alpha '$, $\varepsilon$, $\Lambda$, and $\lambda$.  We can now
integrate equation \appthree\ with respect to $\log ( \varepsilon )$
to obtain ${\tilde Z}^{(0)}$, and insert this quantity in equation
\appone .  Dividing out the volume $\Vol (\Omega ) \propto \log
(\varepsilon )$, we see that the coefficient of the integral of $R$
is independent of $\varepsilon$, $\Lambda$, and $\lambda$.  For this
term, the integral over the superconformal parameters cancels
Vol(SC), and we arrive at equation \Znotagain .

\ack

We would like to thank S.~Shenker and R.~Kallosh for helpful
discussions.  This work is supported in part by National Science
Foundation grant PHY89-17438.  J.~U. is supported in part by a
National Science Foundation Graduate Fellowship.

\refout
\end